\documentclass[acmlarge]{acmart}
\makeatletter
\newcommand{\confshort}{\acmConference@shortname}
\newcommand{\conffull}{\acmConference@name}
\newcommand{\confdate}{\acmConference@date}
\newcommand{\confloc}{\acmConference@venue}
\AtBeginDocument{
  \fancypagestyle{firstpagestyle}{
    \fancyhead{}%
    \fancyfoot[C]{}%
  }
  \fancyhf{}
  \fancyhead[LO]{\@headfootfont\shorttitle}%
  \fancyhead[RE]{\@headfootfont\@shortauthors}%
  \fancyhead[LE]{\@headfootfont\footnotesize \confshort, \confdate, \confloc}%
  \fancyhead[RO]{\@headfootfont\footnotesize \confshort, \confdate, \confloc}%
  \fancyfoot[C]{}%
}
\makeatother
\acmBooktitle{\conffull\@ (\confshort), \confdate, \confloc}

\usepackage{url}
\usepackage{tabularx}
\usepackage{longtable}
\usepackage{booktabs} 
\usepackage[table]{xcolor}
\usepackage{array}
\usepackage[utf8]{inputenc}
\usepackage[T1]{fontenc}
\usepackage{pdflscape}
\usepackage{xltabular}

\copyrightyear{2026}
\acmYear{2026}
\setcopyright{cc}
\setcctype{by}
\acmConference[FAccT '26]{The 2026 ACM Conference on Fairness, Accountability, and Transparency}{June 25--28, 2026}{Montreal, QC, Canada}
\acmBooktitle{The 2026 ACM Conference on Fairness, Accountability, and Transparency (FAccT '26), June 25--28, 2026, Montreal, QC, Canada}
\acmDOI{10.1145/3805689.3806735}
\acmISBN{979-8-4007-2596-8/2026/06}

\begin{document}

\title[From Vulnerable Data Subjects to Vulnerabilizing Data Practices]{From Vulnerable Data Subjects to Vulnerabilizing Data Practices:
Navigating the Protection Paradox in AI-Based Analyses of Platformized Lives}\renewcommand{\shortauthors}{}

\author{Delfina S. Martinez Pandiani}
\email{d.s.martinezpandiani@uva.nl}
\orcid{0000-0003-2392-6300}
\affiliation{%
  \institution{University of Amsterdam}
  \city{Amsterdam}
  \country{The Netherlands}}

\author{Ella Streefkerk}
\orcid{0009-0000-3312-2192}
\affiliation{%
  \institution{Goethe University Frankfurt}
  \city{Frankfurt}
  \country{Germany}}
\email{streefkerk@c3s.uni-frankfurt.de}

\author{Laurens Naudts}
\orcid{0000-0002-5777-1450}
\affiliation{%
  \institution{University of Amsterdam}
  \city{Amsterdam}
  \country{The Netherlands}}
\affiliation{%
  \institution{KU Leuven}
  \city{Leuven}
  \country{Belgium}
}
\email{l.p.a.naudts@uva.nl}

\author{Paula Helm}
\orcid{0000-0002-2719-9721}
\affiliation{%
  \institution{Goethe University Frankfurt}
  \city{Frankfurt}
  \country{Germany}}
\email{helm@c3s.uni-frankfurt.de}

\renewcommand{\shortauthors}{Martinez Pandiani et al. 2026}

\begin{abstract}

This paper traces a conceptual shift from understanding vulnerability as a static, essentialized property of data subjects to examining how it is actively enacted through data practices. Unlike reflexive ethical frameworks focused on missing or counter-data, we address the condition of \textit{abundance} inherent to platformized life—a context where a near inexhaustible mass of data points already exists, shifting the ethical challenge to the researcher’s choices in operating upon this existing mass. We argue that the ethical integrity of data science depends not just on who is studied, but on how technical pipelines transform ``vulnerable" individuals into data subjects whose vulnerability can be further precarized.
We develop this argument through an AI for Social Good (AI4SG) case: a journalist’s request to use computer vision to quantify child presence in monetized YouTube `family vlogs' for regulatory advocacy. This case reveals a ``protection paradox'': how data-driven efforts to protect vulnerable subjects can inadvertently impose new forms of computational exposure, reductionism, and extraction. Using this request as a point of departure, we perform a methodological deconstruction of the AI pipeline to show how granular technical decisions are ethically constitutive. 
We contribute a reflexive ethics protocol that translates these insights into a reflexive roadmap for research ethics surrounding platformized data subjects. Organized around four critical junctures—dataset design, operationalization, inference, and dissemination—the protocol identifies technical questions and ethical tensions where well-intentioned work can slide into renewed extraction or exposure. For every decision point, the protocol offers specific prompts to navigate four cross-cutting vulnerabilizing factors: exposure, monetization, narrative fixing, and algorithmic optimization.
Rather than uncritically embracing AI4SG or dismissing it as purely technosolutionist, we argue for a program of reflexive practice—mirrored in the substantive requirements of European data protection governance—that treats data research as ``world-making" work. This approach moves the researcher from a passive observer to an active agent, inviting methodological reflection to interrogate the shifting boundary between protective visibility and predatory exposure.
\end{abstract}



\maketitle

\textit{``A vulnerability must be perceived and recognized in order to come into play in an ethical encounter, and there is no guarantee that this will happen 
[...] vulnerability is fundamentally dependent on existing norms of recognition if it is to be attributed to any human subject.''}
-- J. Butler, \textit{Precarious Life: The Powers of Mourning and Violence} (2004, p.~43)

\section{Introduction}

This paper interrogates a troubling paradox in data-intensive research: how the very practices designed to protect `vulnerable' data subjects can inadvertently impose new forms of computational reductionism, exposure, surveillance, and control. Our point of departure is a concrete request. A journalist approached our lead author, asking them to employ AI tools—specifically computer vision-based facial recognition, emotion detection, and sensitive content recognition—to quantify the presence of children in monetized YouTube family vlogs. Their goal was to empirically substantiate a push for stricter regulation of the monetization of these vulnerable subjects. On its surface, this request appears to sit squarely within the established imaginary of “AI for Social Good” (AI4SG) \cite{Kanza_2020_AI4Good_WorkshopSummary}: through the use of AI, our lead author would provide empirical evidence of the monetization of children’s intimate lives, along with its downstream effects, including the circulation of sexualized attention and grooming vectors. We would equip regulators with numbers and patterns. We would leverage technical skill and bend it toward protection. This is the familiar promise that data science can help us “see” wrongdoing at scale, and thereby compel remedy and restore justice \cite{DIgnazio_2023_toolkit, D’Ignazio_2024}.

We argue that beginning from this request, which both enticed our lead author and left them feeling uncertain about fulfilling these requests at all, helps us think through a central tension in contemporary AI4SG practice, which D'Ignazo and Klein in their Data Feminism program have described as the \textit{paradox of exposure} \cite{DIgnazio_Klein_2020_DataFeminism}: the very same systems that promise to surface harm and inform responsive policy also reproduce, and in some cases intensify, the conditions that make those harms durable. To quantify children in vlogs, one must find, extract, and render them into machine-readable form. One must detect faces, infer affect, track presence across videos and platforms, and attach labels to bodies and speech. One must produce datasets that consolidate, stabilize, and recirculate children’s appearances, gestures, moods, and family dynamics. In the process, children become legible as objects of governance by first becoming objects of computation. Protection \textit{from} capture is enacted through another act \textit{of} capture. Additionally, the question can be raised of whether reliance on intrusive, data-driven technologies in non-research contexts legitimises their use within research contexts. In other words, how does research relate to the broader deployment of technologies within society?

Rather than treating this as a familiar trade-off between privacy and accountability, we suggest it discloses a deeper paradox researchers too must face: those with technical and institutional power claim the authority to intervene on behalf of those depicted as vulnerable, worthy of special protection, while the intervention itself extends the reach of extraction, surveillance, and categorization that made proactive measures of protection seem necessary in the first place. The protection paradox emerges because, as philosopher Judith Butler (\citeyear[p.43]{butler_precarious_2004}) notes, vulnerability does not automatically trigger an ethical response; it remains ethically inert until filtered through specific regimes of perception and recognition. This recognition is increasingly mediated by datafication. For a harm, a risk, or a right to be recognized—and thus become an issue worthy of intervention—it must first be rendered into a datafied representation. Consequently, the very act of recognizing vulnerability—the prerequisite for any ethical encounter or protective intervention—has become dependent on the same computational infrastructures of extraction and categorization that lead to the need for protection. Here, we follow Butler’s approach where vulnerability is treated not merely as a universal condition of life \cite{cole_all_2016}, but also as an effect of power, infrastructures, and incentives that make some lives more easily knowable, tradable, narratable, and actionable by others, thereby contributing to what Butler describes as ``precarization" \cite{butler_precarious_2004}. This precarization depends on many factors and can apply to many different groups of people. These dynamics have been exposed already in the context of the AI for Development (AI4D) discourse, where labels such as “vulnerable populations” position certain groups as recipients of rescue by analytic expertise, thereby instead of contributing to their empowerment, reproducing dependency and epistemic marginalization \cite{Madianou_2024_Technocolonialism, Schelenz_Pawelec_2022_ICT4Dcritique, helm2023diversity}. While our study focuses on families operating under EU law—a vastly different legal and social context—we argue that this paradox more generally applies when data subjects’ lives become \textit{platformized}. By this, we mean lives that are made public and rendered as objects of monetization within the attention economy \cite{erin_duffy_politics_2024}. In our exemplary case, this concerns children featured in European family vlogs. 

Critically, this case presents a research gap that is at once practical and conceptual. The case is significant in its own right, but also because, unlike contexts in which data feminism calls for making injustice visible by filling in \textit{missing} data \cite{dignazio_counting_2024}, platformized life is defined by a condition of data abundance where billions of data points already exist. Here, the ethical challenge is not one of visibility through data collection, but of the reflexive responsibility involved in data practice. The tension lies in how we, as data scientists and researchers, operate upon this existing abundance.
In this work, we advance the view that data practices must be understood as potentially \textit{vulnerabilizing} (precarizing vulnerabilities), especially within interventions that aim to “do good,” “protect,” or “serve” vulnerable subjects through data science and AI. We aim to foster reflexivity for data-driven analysis, specifically in the context of platformized lives by situating these practices along the lines of the “protection paradox.” This entails a reflexive engagement with the situated routines and data infrastructures through which visibility and recognition are produced, circulated, and locked in as evidence. These practices include reflection on how datasets are designed and assembled, what kinds of labels are imposed, which proxies and inferences are treated as legitimate, and how findings are framed and disseminated. While this dynamic involves a broad network of actors—including platforms, parents, and regulators—this paper focuses specifically on the role of the data scientists. We aim to provide the conceptual clarity to understand how vulnerability is algorithmically produced and under what socio-technical routines and conditions these interventions are staged. 

In the sections that follow, we elaborate this argument in five moves. Section \ref{sec:ai4sg} situates the protection paradox of AI4SG initiatives within debates in data feminism and restorative justice. Section \ref{sec:vulnerability} mobilizes insights from political philosophy on the moral status of vulnerability to demonstrate how protective intentions in AI4SG can generate new forms of exposure and precarity. 
Section \ref{sec:fv} analyzes our case study of child presence in family vlogs, organized around four stages of the data science pipeline: \textit{dataset design, operationalization, inference,} and \textit{dissemination}. Rather than treating ethical tensions as case-specific problems, we show how they reveal general mechanisms through which vulnerability is precarized in AI research on platformized lives. 
Section \ref{sec:vdp} synthesizes these stages to identify four cross-cutting vulnerabilizing factors—exposure, monetization, narrative fixing (a form of epistemic foreclosure where a subject’s identity is externally scripted and stabilized), 
and algorithmic optimization—and introduces a \textbf{reflexive protocol for data scientists} (Appendix \ref{appendix}) to recognize and negotiate critical technical decision points. Finally, Section \ref{sec:law} grounds our discussion in European data protection legislation, demonstrating how the GDPR can be interpreted to mandate a vulnerability-aware approach wherein legal standards and reflexive practices are mutually reinforcing: the law provides the institutional mandate for care, while our protocol provides a technical orientation for its execution. Sections \ref{sec:limitations} and \ref{sec:conclusion} address the limitations of our case-study approach and conclude by outlining broader implications for the future of AI4SG research and the governance of platformized lives.

\section{The AI4SG Protection Paradox}
\label{sec:ai4sg}

AI for Social Good (AI4SG) has emerged as a capacious label under which efforts to align AI-driven inference with public value are gathered. It couples consequentialist ambitions, such as welfare gains and access to resources, with deontic constraints, rights, autonomy, justice, as well as epistemic and procedural virtues such as transparency and accountability. Within this setting, the AI4People framework, advanced by Floridi and collaborators, codifies beneficence, non-maleficence, autonomy, justice, and explicability, urging institutions to build ethics “upstream” of deployment so that legitimacy inheres in socio-technical arrangements rather than in post hoc compliance \cite{Floridi_2018_AI4People, taddeo_how_2018}. While this framework is aimed at ``the good AI society", it is also acknowledged that ``the good society" does not exist in any absolute sense. Instead, comparative policy analysis exposes that what counts as “good” is politically mediated: US, EU, and UK traditions weigh innovation, rights, and risks very differently and translate shared principles into divergent legal and standardization pathways \cite{Cath_2018_good_society}.

Institutionally, AI4SG has been stabilized through platforms and research venues that are multi-stakeholder and promise to connect technical agendas to public outcomes. UN-adjacent forums emphasize capacity building; workshops and hubs such as the NeurIPS 2019 Joint Workshop on AI for Social Good and the AI for Social Good repository \cite{ITU_nd_AIFG_about} foreground problem selection, limitation statements, and policy coupling. Consulting-style mappings highlight cases aligned with the Sustainable Development Goals, pairing technosolutionist optimism with caveats about data access, talent, and “last mile” implementation \cite{Chui_2018_MGI_AISG, NeurIPS_2019_AISG_CFP}. Syntheses of scientific practice, however, continue to flag unresolved issues of bias, provenance, reproducibility, and incentive structures, warning that techno-optimism cannot substitute reforms of infrastructures, business models, and institutions \cite{Kanza_2020_AI4Good_WorkshopSummary}. Critical perspectives therefore press AI4SG to reckon with power, positionality, and asymmetric institutional contexts. Hardcastle, for instance, argues that AI4SG discourse abstracts away the situated and uneven character of platformized publics, and instead calls for a socially embedded reframing that centers contestation, institutional design, and the distribution of benefits and burdens \cite{Hardcastle_2024_RethinkingAIFG}. Other commentators suggest that AI4SG risks devolving into technosolutionist public relations: rather than supporting sustainable change, local capacity building, and collective empowerment, it can end up disenfranchising precisely those subjects in whose name “the good” is invoked \cite{Hardcastle_2024_RethinkingAIFG, Moorosi_2023_AIforWhom, Birhane_2025_false_promise, Nutas_2024_AISolutionism_GAIA, Zuger_Asghari_2024_PublicInterestIntro}.

We align with the skepticism and look to alternative approaches that begin from power and inequality rather than abstract ethical desiderata. The Data Feminism proposal is exemplary here \cite{DIgnazio_Klein_2020_DataFeminism}. Instead of foregrounding beneficence in the abstract, it treats reflexivity around structural inequality as a first-order design variable. Data feminism proposes principles that start from uneven distributions of harm and privilege, asking data work to challenge power by making labor visible, embracing plural ways of knowing, and designing explicitly for equity rather than an ideal of neutrality. A companion program of restorative and transformative data science operationalizes these ideas through community-led problem definition, data minimization and refusal where collection would reproduce harm, participatory risk assessment, and life-cycle stewardship that includes repair, maintenance, exit, and handover as ethical phases of a project \cite{dignazio_counting_2024}. Crucially, such approaches insist on the ambivalence of visibility: there are situations in which invisibility is a form of protection, in which not collecting data is the better choice, and in which listening to data subjects’ own perspectives must precede any assumption about what they “need” or “want” from data work \cite{Benjamin_2019}.

We hence take abolitionist critiques seriously and do not assume that further datafication is inherently desirable. In many cases, testimony about harm should already be sufficient to justify refusal, restriction, or abolition. However, under AI4SG and current EU regulatory logics, datafication often remains a condition of legibility, legitimacy, and institutional response. Our intervention is therefore pragmatic: not to uncritically endorse these arrangements, but to make vulnerabilizing practices visible within them, while supporting refusal and constraint where appropriate. Our theory of change is to render harms actionable inside existing governance pipelines so that accountability demands become harder to dismiss under present institutional conditions.

This paper enacts this theory of change through a concrete intervention. Responding to the aforementioned journalists’ requests to use computer vision to quantify children’s presence in family vlogs, we examine the ethical tensions that arise when such projects move from proposal to practice.  In doing so, we identify a research gap that emerges specifically at the intersection of AI-driven ``social good" and the hyper-visibility of platformized life. While researchers are readily equipped with AI tools that can surface harmful patterns and inform differentiated policy, these same tools tend to re-enact extraction, extend surveillance, and deepen asymmetries between those who wield them and those being ``protected." Crucially, our case identifies a limit in current reflexive frameworks, which often focus on challenging power by making injustice visible through ``counter-data" collection \cite{D’Ignazio_2024}. In contrast, our study concerns hyper-visible mass phenomena in which billions of data points already exist. The ethical challenge here is not one of visibility, but of navigation and re-exposure. The difference lies less in the data points themselves than in \textit{how we look} at and analyze them, \textit{how we reflect} on our own practices, and \textit{how far} these practices actually contribute to achieving the ends they were intended to serve.

\section{Vulnerability: A Universal Condition in Need of Recognition}
\label{sec:vulnerability}

In legal and feminist theory, vulnerability is regularly defined as \textit{potentiality of harm} \cite{gilson_ethics_2014}: a condition of openness to being affected or wounded within unequal distributions of power, care, and recognition. In normative and legal frameworks, this potentiality underwrites protective measures such as data governance regimes. As Fineman (\citeyear{fineman_vulnerable_2008}) argues, vulnerability is a universal condition: all living beings possess interests \cite{martin_resolving_2014-1}, and it is precisely their capacity to be harmed that grounds the need for social and institutional safeguards. 

While all individuals are vulnerable, some experience heightened precariousness not only because of ontological fragility (as in the case of children’s smallness or developmental dependence), but also because of political, economic, and technological structures, social prejudices, and historical inequalities \cite{mackenzie_introduction_2013}. Recognizing that vulnerability can more readily become precarious for certain groups is politically significant \cite{peroni_vulnerable_2013}. This is why legal scholars have suggested translating these differences into layered categories, such as ``vulnerable,'' ``especially vulnerable,'' ``particularly vulnerable'', or even ``extremely vulnerable" subjects \cite{luna_elucidating_2009}. More concretely, information law and consumer protection regimes already distinguish between the average user and those in need of heightened safeguards (children, the elderly, the incarcerated, socio-politically marginalized users) \cite{malgieri_vulnerable_2020}.
From an ethical perspective, the notion of vulnerability is, in effect, treated as a heuristic for the identification of social injustice \cite{fineman_vulnerable_2008}, which members of \textit{particularly} vulnerable groups are\textit{ more} likely to experience \cite[p. 1064-1065]{peroni_vulnerable_2013}. In this context, ascriptive identity traits, such as a person’s sexuality, mental faculty, or migrant status,\footnote{See, for example \cite{ECHR_DH2007, ECHR_MSS2011}} act as signifiers for the (mutually) reinforcing social processes that underlie and engender, yet also beget, future vulnerabilities. This hybrid character of vulnerability---all people are inherently vulnerable, while for some their inherent vulnerability becomes more precarious as a result of discrimination, surveillance, hyper visibility, sexualization, etc.---is reflected in such formulations. However, the very act of codifying vulnerability in law encourages its proceduralization: it becomes something to be operationalized through criteria, thresholds, and checklists, thus, again, reducing its complexity.

Research ethics inherits and amplifies these dynamics. Institutional Review Boards (IRBs) and research ethics committees rely mostly on biomedical and psychological notions of ``vulnerable populations,'' which are then transposed into data-intensive contexts. In EU Horizon funding templates, for example, vulnerability often appears as a box to be ticked: researchers are asked whether they work with ``vulnerable participants'' and, if so, to outline additional safeguards. Vulnerability is thereby rendered as an attribute of certain pre-defined groups, and as a potential \emph{obstacle} to research: a problem to be managed efficiently so that the project may proceed.

Critical scholars have warned that such categorical attributions, while motivated by protection, carry risks \cite[p. 1064-1065]{peroni_vulnerable_2013}. First, drawing sharp distinctions between groups deemed more or less deserving of protection can silence or marginalize those whose experiences fall outside dominant narratives of vulnerability \cite{butler_precarious_2004, butler_vulnerability_2016}. Second, externally assigning vulnerability may reinforce paternalistic dynamics and undermine the autonomy and agency of those being classified \cite{cole_all_2016}. Such an approach may thereby enhance disenfranchising dynamics of AI4SG research, outlined above. The figure of ``the vulnerable data subject'' risks obscuring the granular ways in which concrete data practices expose some people, and in some situations, more than others.

Children are often treated as the exception that confirms the rule. Their need for protection is widely acknowledged and typically grounded in physical and cognitive immaturity, including smallness and developmental dependence \cite{engle_child_1996}. In legal doctrine, this is formalized through categories such as “especially vulnerable minors.” Yet even within childhood, vulnerability is far from uniform, varying with socio-political positioning, degrees of privilege, and the extent to which decisions by guardians, platforms, or institutions shape children’s lives. Not all children experience exposure, manipulation, or monetization in the same way or to the same degree; for some, the potential for harm is much more likely to be realized. Binary distinctions between “average” and “vulnerable” users therefore fail to capture the temporal, contextual, and relational emergence of vulnerability across digital environments, and the platformization of private lives has made these dynamics sharply visible \cite{erin_duffy_politics_2024}.

At the same time, vulnerability is not only a condition of subjects but also an effect of situations, practices, and infrastructures—and one that becomes political only when recognized. As Butler argues, “a vulnerability must be perceived and recognized in order to come into play in an ethical encounter,” such that “vulnerability is fundamentally dependent on existing norms of recognition if it is to be attributed to any human subject” \cite{butler_precarious_2004}. Recognition, however, is never neutral. Vulnerabilities can remain unnoticed—rendered invisible, illegible, or unspeakable—and even when acknowledged, recognition transforms what vulnerability is taken to mean and how it is structured. Norms of recognition thus exert significant power in shaping which harms count, which subjects are deemed at risk, and which responses appear appropriate. Increasingly, these norms are mediated through data-driven infrastructures in which vulnerability must be demonstrated, measured, or computed in order to be seen.

These tensions surface acutely in data-intensive research and AI4SG projects. Data science and AI systems are powerful tools, and those who deploy them wield considerable influence over how people are rendered visible, categorizable, and actionable. In this context, the proceduralization of vulnerability through legal or research-ethics templates that ask \emph{who} is vulnerable can obscure the more substantive question of \emph{how} particular data practices themselves produce or amplify vulnerability. When the promise of AI4SG is to protect people online, detect abuse, or mitigate harms, an ethics checklist organized around binary classifications of vulnerability is insufficient. Starting from the case study, in the following section we inductively map decision points in AI pipelines where specific data practices contribute to the precarization of vulnerability. This mapping supports a more nuanced and reflexive engagement with how vulnerability is shaped through data practices.

\section{Case Study: Data-driven Protection of Children in European Family Vlogging}
\label{sec:fv}

The data subjects at the center of this case study are children prominently featured in European family vlogs (FVs). These publicly available, highly visible, and monetized videos are cultural artefacts shaped by platform logics of visibility and extraction \cite{Dijck_2018} in which children’s everyday lives become key drivers of engagement and revenue \cite{ErinDuffy_Ononye_Sawey_2024}. FV transforms domestic life into a market-oriented performance \cite{abrams2023}, often centering children in affectively charged scenes such as illness, conflict, or loss \cite{Livingstone_2025, Sefton-Green_Mannell_Erstad_2025}. A monetized subcategory of sharenting, FV amplifies tensions between parental rights and children’s rights to privacy and digital autonomy \cite{abrams2023, Steinberg_2017}. Public discourse increasingly treats these children as vulnerable, situating FV within a broader landscape of platform governance and regulation (e.g., \cite{weiss2023nypost, valentinodevries2024nyt, miller2025nyt, smidt2017buzzfeed}). In this environment, policymakers and journalists seek empirical, computational evidence of children’s exposure and risk. 

The lead author was approached by a journalist to analyze one year of content from an extremely successful Dutch family vlog channel, with the purpose of quantifying (1) how often the family children appear on screen (as evidence for potential labor exploitation), (2) which emotions they display (to highlight psychological exposure), and (3) the frequency of `sensitive' scenes involving partial nudity (to demonstrate risks of sexualization and downstream harm). 
These are quintessential AI4SG requests: data science used to protect and regulate. 
A broader technical team (of which the lead author of this article is a member) began considering these requests, and designing a data pipeline for potentially fulfilling them. 
Given the scale and visual nature of the content, these protective aims require deployment of AI, and specifically of computer vision (CV) methods, through tasks such as face detection or recognition, facial emotion analysis, and NSFW (Not Safe For Work) or sensitive scene classification that transform human bodies, expressions, and domestic contexts into machine-readable data. Drawing on large-scale image corpora scraped from platforms, these systems embed social norms into technical pipelines and ultimately recirculate these norms back into social and regulatory life \cite{crawford2021excavating, birhane2021large, barocas2023fairness}. 
The aforementioned technologies have controversial histories and use-cases that mandate both skepticism and caution \cite{kalluri2023surveillance, kalluri2025computer, crawford2021atlas, mcstay2023automating}, especially when they are used to detect abstract concepts with nuanced cultural meanings \cite{martinez2024wicked} such as emotions, social values and loosely defined umbrella terms like toxicity \cite{pandiani2025toxic}. For instance, affective computing technologies have been questioned for their pseudo-scientific foundations and purported societal benefits \cite{mcstay2023automating, crawford2021atlas}. 

The reflective exercise we endorse demands from data researchers serious engagement with these criticisms, not only within isolated research settings (can a particular technology scientifically deliver what they promise in a manner that does not significantly harm data subjects?), but also on how research involving these technologies may contribute to their broader societal adoption (can affective computing in research perpetuate a worldview that approaches emotions primarily as measurable states of being?). An overarching question underlying our framework, therefore, is: should we engage with intrusive technologies in research at all? In certain cases, and regardless of the conditions research may put in place, the answer should be \textit{no}.

Confronted with the ethical and legal fraughtness of the case at various technical design levels, and with the scarcity of guidelines for ethically operating upon abundant platformized data,
the lead author invited the other three co-authors of the present paper to systematically reflect on the emerging pipeline and on how protective intentions can themselves generate vulnerabilizing data practices. Following recent calls from HCI to use autoethnographic case studies to develop more generalizable protocols (e.g., scaffolded autoethnography \cite{ajmani2025movingtowards}),
we methodologically adopt a first-person, reflexive analytic stance on this ongoing applied project. 

\subsection{Data Practices in Context: Four Pipeline Stages}

We structure our analysis around four recurring stages in many data-science (and increasingly AI-led) pipelines: 
(A) dataset design, (B) operationalization, (C) inference and evaluation, and (D) dissemination. At each stage, already present vulnerabilities can be amplified or transformed through technical choices, and the stages themselves are interdependent.
At the time of writing, technical implementation of the AI pipeline for the FV case study has advanced through dataset design and operationalization, being intentionally halted prior to inference, evaluation, and dissemination, in order to allow for the space needed  for systematic reflection. 
The case has therefore not culminated in a fully implemented AI system. We treat this suspended pipeline—including already-enacted and not-yet-enacted steps—as an object of inquiry in its own right, deconstructing it into a series of decision points rather than reporting on completed outcomes.

The following sections describe the case-specific tensions at each stage, showing how the push for “evidence” collided with technical realities. While grounded in the FV case, the named techniques exemplify broader families of data-intensive methods. Table~\ref{tab:decision-points} abstracts from these particulars to show how each technical decision point surfaces generalizable ethical tensions and mechanisms of \textit{precarized vulnerability}. Appendix~\ref{appendix} provides a richer version of the table, including reflexive questions and potential responses. This vulnerability-aware protocol is thus an inductively derived framework that formalizes the frictions and refusals encountered during our process. Understood as a form of analytic generalization, this protocol is intended as a transferable tool for other researchers and practitioners to begin from and adapt when reflecting on their own data-driven AI4SG projects.

\subsection{Stage 1: Dataset Design}
Dataset design is an ethically constitutive phase in which decisions actively shape who becomes visible, analyzable, and subject to protection or harm. In the context of analyzing FV, these decisions determine what kinds of harm can be evidenced, and also how exposure and scrutiny are distributed across children, families, and populations.

\textbf{Scale.} The journalist’s request was targeted at a single family vlog channel. In tension with this, the ethics board encouraged broader sampling, foregrounding a fundamental trade-off: narrow sampling (one channel/household) concentrates scrutiny and cumulative exposure on a small number of identifiable children, whereas broader sampling (e.g. all content listed as family vlogging in the Netherlands) diffuses this fixation but expands the population subjected to algorithmic screening and surveillance. 
To preserve contextual interpretability while limiting hyper-fixation on one family, implementation proceeded with a sample of 4 family vlogging channels.

\textbf{Subject selection criteria.} Selecting channels based on controversy or high engagement—incentivized for regulatory relevance— mirrors platform reward logics. This can precarize the vulnerability of children by hyper-scrutinizing those already heavily exposed, treating ``attention-optimized" extremes as genre-wide baselines. For sensitive imagery detection, for example, engagement-driven sampling may overrepresent families already circulating in sexualized attention economies, skewing risk prevalence while hyper-scrutinizing a targeted subset. 
Conversely, purely random sampling would obscure how platform architectures systematically reward particular kinds of content. 
Given the platform regulatory purpose of the AI intervention, channels were selected based on engagement metrics, foregrounding content that platforms already elevate and financially reward.

\textbf{Comparability and aggregation.} Deciding whether subjects are aggregable units or distinct cases involves a trade-off between contextual flattening and pattern detection. 
In emotion recognition, for example, aggregation across videos presumes comparability, potentially misinterpreting performative affect in ``prank" videos. 
In sensitive content detection, disaggregation risks framing systemic issues (e.g. patterns of nudity across channels) as idiosyncratic. 
Aggregation for presence estimation was performed at the video and channel levels, while tracking content type and explicitly acknowledging limitations of this design choice.

\textbf{Data instance granularity.} Deciding the resolution of data instances defines the intensity of capture. 
Fine-grained analysis (e.g. individual video frames) offers higher precision and reliability (especially for presence estimation), 
but it also increases inferential density: subjecting children to thousands more discrete computational judgments. 
More coarse analysis (e.g. thumbnails) may miss important instances relevant to investigative claims such as those
regarding child labor. 
Implementation proceeded with frame-based sampling for presence estimation.

\textbf{(Public) access and reuse.} Relying on ``public accessibility" as an ethical proxy for consent collapses the gap between availability and permissibility. 
This is fraught for children who can neither consent to nor exit the ``extended data lives" research creates. 
In our case, inclusion may legitimize the repeated inspection of domestic spaces.
For affective tasks, reuse creates secondary data traces outliving the original context; for sensitive imagery, it moves intimate material from ephemeral platform environments into (potentially more permanent) research infrastructures. 
Implementation proceeded with use of public content of only 4 FV channels within the scope of evidentiary aims, anonymized wherever possible, and set to be deleted immediately after gathering of results.

\subsection{Stage 2: Operationalization}

Operationalization—the process of turning abstract research questions into measurable computational objects—is ethically constitutive. In our case study, this is where (performative) family dynamics are flattened into discrete labels, creating the ground truths upon which all subsequent evidentiary claims are built.

\textbf{Unit of analysis and localization.} Even after determining dataset granularity, the resolution at which each analysis is localized defines the subject's visibility. For presence estimation, global frame-level detection (is \textit{any} child present?) avoids the creation of identifiable biometric databases but obscures individual labor disparities, such as between siblings. Localized identification (detecting and tracking specific children) provides evidentiary weight for labor regulation but requires generating and storing sub-datasets of minors' faces. 
This choice further dictates whether ``presence" requires a facial view or a bodily silhouette; for regulatory labor claims, any presence (even looking away) may be sufficient, but for affective tasks, clear facial views may be needed. 
This choice determines whether we produce person-specific profiles that subject the children to intensified computational tracking. 
For presence estimation, implementation proceeded with frame-level person detection, upon which a global analysis (are any of the people children?) of child presence is performed, 
as creation of identifiable biometric archives of minors was not deemed justifiable for the purpose of labor claims.

\textbf{Taxometric architecture.} The logical structure of a classification task—whether binary, multi-label, or continuous—defines the ``thickness" of automated scrutiny. 
For emotion recognition, we grappled with binary classification (e.g., positive/negative), multi-class single-label (one from discrete buckets), or continuous scoring (e.g., a 0–1 intensity scale). 
While binary models offer technical simplicity, they can impose an extreme categorical reductionism (e.g. transform a playful, transient scream into a durable, machine-readable ``high-intensity distress" discrete metric, leaving no room for performative ambiguity). 
Conversely, continuous or multi-label architectures can preserve more ``nuance" but increase inferential density. 
For presence estimation, implementation proceeded as a binary (presence or lack thereof) task.

\textbf{Taxonomic target definitions.} Selecting a taxonomy is an act of normative imposition. 
In our case, using certain emotion categories or specific nudity thresholds can reflect cultural and normative assumptions (especially adult-centric and WEIRD—Western, Educated, Industrialized, Rich, Democratic \cite{Henrich2010Weirdest}—ones), assuming bodily norms and internal states as universally legible, and misinterpreting playful, performative, or neurodivergent expressions. 
Labels can also carry heavy moral connotations: reducing complex affects to singular labels like ``sad" or ``fearful" can inadvertently script a narrative of trauma or neglect. 
In the operationalization of sensitivity, defining ``nudity" through fixed skin-exposure percentages might misinterpret ordinary caregiving as inherently suspect, contributing to the stigmatization or objectification of children's bodies we intended to protect, while using scene-based tags (e.g., ``bathing," ``medical") can subject families to very invasive domestic cataloging. 
For presence estimation, implementation proceeded with a binary target: child present/child non-present.

\subsection{Stage 3: Inference and Evaluation}

Inference and evaluation are the technical execution of a research plan, but they are also the phase where probabilistic outputs are converted into definitive ``evidence." This stage introduces a tension between the pressure for clear regulatory results and the inherent instability of algorithmic interpretation. Ultimately, the researcher’s choice of implementation and the interpretation of error rates dictate whose lives are rendered computationally legible and whose are distorted through algorithmic misfire.

\textbf{Infrastructure and data sovereignty.} Deciding where to run inference—locally in secure servers or via third-party APIs (e.g., OpenAI, AWS)—constitutes a significant privacy threshold. Using hosted models, including Large Language Models (LLMs), necessitates transmitting sensitive, often unconsented data of minors to Big Tech infrastructure. This forces a tension between analytical power and data sovereignty: state-of-the-art reasoning comes at the cost of exposing subjects to secondary data gazes and permanent storage within proprietary systems.
All inference will proceed only with local models in local servers.

\textbf{Implementation and model selection.} Selecting a model implementation, including through frozen pre-trained encoders or task-specific fine-tuning, can mask demographic and cultural variance. In our case, applying models trained on adult actors for emotion recognition to children’s domestic vlogs forces idiosyncratic behaviors through a narrow representational sieve. For instance, technologies claiming they can detect emotions based on visual cues have been shown to misinterpret the faces of people, including children, of color as ``aggressive". As such, relying on off-the-shelf models can scale systemic biases, rendering marginalized individuals more susceptible to automated misclassification and visibility distortions.
For presence estimation, inference will proceed with a Vision Transformer finetuned to classify images of human faces into `minor' or `adult'.\footnote{\url{https://huggingface.co/Civitai/age-vit}}

\textbf{Stochasticity.} The integration of non-deterministic models (like LLMs) into the inference pipeline introduces stochasticity. Unlike traditional CV models with stable outputs, LLM-based reasoning can produce varying interpretations of the same domestic scene based on prompt phrasing or temperature settings. This raises reliability concerns: if interpretations of ``distress" fluctuate with arbitrary prompt changes, the research risks producing inconsistent social verdicts that depend more on the stochastic nature of the model than the lived reality of the child. 
All inference will proceed with model architectures and inference strategies that are deterministic.

\textbf{Thresholding.} The translation of a model’s probabilistic output into a stable ``detection" relies on setting thresholds that carry distinct ethical consequences. In sensitive imagery detection, the researcher must decide the ``bar" for what counts as a fact (e.g., $p > 0.9$). Models trained on adult norms may fail to flag ordinary but sensitive domestic moments—such as potty training or bathing—as ``sensitive," even when they involve partially nude minors. These false negatives can erase the harms and increased vulnerabilization that the research aims to make visible. Conversely, a lower threshold may capture these moments but multiply false positives that pathologize benign caregiving. Researchers must further navigate the legal and personal boundaries of investigating intimate scenes, as the act of ``detecting" can itself become an invasive crossing of domestic privacy.
For presence estimation, a threshold of 0.9 is first used to detect whether the detected object is a person. There is no explicit threshold for whether that person is a child or adult: instead, the class (adult or minor) with the highest probability is selected.

\textbf{Evaluation metrics and contextual relevance.} Validating results through aggregate metrics like Accuracy or F1-scores can mask disparate impacts and reinforce problematic social narratives. For example, an emotion recognition model might achieve high accuracy while consistently failing on neurodivergent children or specific socioeconomic domestic settings. Relying on technical performance alone creates a validation trap, where the model's truth overrides the subject's lived experience. Misclassification can exaggerate harm, potentially over-pathologizing specific family structures, or erase it, such as by missing signs of genuine discomfort. This forces us to interrogate whether technical limitations can reinforce existing political narratives about the families or the phenomenon of vlogging itself.
For presence estimation, on top of reporting the accuracy of the selected off-the-shelf detector, manual auditing of ~5000 frames will be done to ascertain the child detector's performance in the context of the FV dataset.

\subsection{Stage 4: Dissemination}

The dissemination stage is the active curation of a ``secondary life" for the data and results, determining how computational findings move from controlled research environments into peer-reviewed archives, policy briefs, and the public imagination. This surfaces the tension between the researcher’s responsibility to evidence harm and the risk of inadvertently scaling the exposure or stigmatization of the very subjects they sought to protect.

\textbf{Narrative framing.} 
Synthesizing complex model outputs into a coherent narrative about children in FV requires balancing analytical 
humility, based on probabilistic model outputs, with the rhetorical clarity necessary for regulatory change.
Framing a model’s output as definitive proof of a moral or legal state can 
lead to narrative fixing: the stabilization of subjects within certain roles (such as that of the `exploited child'), 
which can even trigger unmerited social or state intervention.
Interpretive closure that overrides the subject's own narrative is particularly fraught when findings 
are translated for news media or policy-makers who may lack the literacy to interpret algorithmic uncertainty.
Presence estimation is to be explicitly reported using qualifying language that acknowledges the limitations of model accuracy and of algorithmic ascription over lived experience, and refraining from reproducing oversimplified narratives.

\textbf{Visualization.} 
Decisions regarding the inclusion of platform content in papers or conference presentations—even when blurred—introduce 
a tension between evidentiary impact and cumulative exposure. 
Circulating screenshots risks further exposing children whose images already circulate widely, 
potentially converting protective aims into new forms of public or economic harm by freezing children into stylized roles that echo the attention economy of vlogging itself.
While `visual proof' (a sensitive domestic scene or a child's moment of distress) may help prove a regulatory claim, 
it permanently archives that moment within the academic record and reinforces exposure to further audiences. 
Researchers must navigate whether including such aestheticization of private life serves to protect minors or merely creates a new, durable site of spectacularization. 
For presence estimation, screenshots of frames are not to be included.

\textbf{Publication \& circulation.} 
The act of publishing creates a digital footprint that outlives the original platform environment, 
giving research (meta)data a ``secondary life" beyond its initial context. 
Once released, findings on FV family life become available for downstream reuse 
by journalists, researchers, companies, and private citizens. 
While generating evidence supports the pursuit of public accountability and transparent research practices, 
it can be weaponized by bad actors including for vigilante harassment (of private citizens against FV parents),
or by platforms and companies (e.g., health insurance) against the data subjects in their future social or clinical lives. 
Additionally, there is a risk that sharing certain results—such as sensitive content trends—might create a roadmap for finding intimate or nudity-prone scenes, 
rendering them hypervisible to new audiences.
In our case, only channel-level aggregated numbers are to be published, and no channels names are to be included.

\begingroup
\footnotesize
\renewcommand{\arraystretch}{1.5}
\begin{longtable}{|p{0.12\textwidth}|p{0.27\textwidth}|p{0.21\textwidth}|p{0.34\textwidth}|}

\caption{Stages and technical decision points recurrent across (AI-led) data science pipelines. The table traces how each technical question and decision point can surface ethical tensions and potential precarization of data subjects' vulnerabilities.} \label{tab:decision-points} \\

\hline
\rowcolor[gray]{0.8}
\textbf{Decision} & \textbf{Technical Question} & \textbf{Ethical Tensions} & \textbf{Potential Vulnerabilizations} \\ \hline
\endfirsthead

\hline
\hline
\rowcolor[gray]{0.8}
\textbf{Decision Point} & \textbf{Technical Question} & \textbf{Ethical Tensions} & \textbf{Potential Vulnerabilizations} \\ \hline
\endhead

\hline
\multicolumn{4}{r}{{Continued on next page...}} \\
\endfoot

\hline
\endlastfoot

\multicolumn{4}{|l|}{\cellcolor[gray]{0.9}\textbf{STAGE A: DATASET DESIGN}} \\ \hline
\textbf{Scale} & How many data subjects are included in the dataset? (e.g. one family, all monetized Dutch family vlogs) & \textbf{Subject-specific fixation vs.\ population-level scrutiny} 
& Hyperfixation and intensified scrutiny of specific subjects; increased traceability and narrative fixing over time.
\\ \hline

\textbf{Selection \newline criteria} & On what grounds are subjects rendered ``eligible''? (e.g. engagement threshold, media scrutiny) & \textbf{Attention mirroring vs. \newline structural opacity} 
& Intensified scrutiny of already-visible subjects; representational fixation in which specific subjects become emblematic cases. 
\\ \hline

\textbf{Comparability \newline \& aggregation} & Under what assumptions are subjects treated as comparable? (e.g. siblings from the same family) & \textbf{Flattening vs. \newline over-individualization} 
& Flattening that obscures differential vulnerability; normalization of certain bodies as baselines for risk. 
\\ \hline

\textbf{Resolution \& granularity} & At what granularity are subjects rendered analyzable? (e.g. thumbnails vs.\ frames) & \textbf{Analytical sensitivity vs. \newline inferential density} 
& Hyper-granular exposure; intensified inferential density through repeated inspection.
\\ \hline

\textbf{Access \& reuse \newline justification} & On what grounds is data deemed legally eligible for inclusion? (e.g. public access, monetization, consent) & \textbf{Legal accessibility vs. \newline ethical permissibility} 
& Extended data life through recontextualization; normalization of platform-defined publicness as ethical consent. 
\\ \hline

\multicolumn{4}{|l|}{\cellcolor[gray]{0.9}\textbf{STAGE B: OPERATIONALIZATION}} \\ \hline
\textbf{Unit of \newline analysis \& \newline localization} & At what resolution is the analysis performed? (e.g. global vs.\ localized) & \textbf{Hyper-identification \newline vs. contextual blur}
& Persistent individual tracking; creation of biometric corpora of minors; labor erasure via de-individualization of background presence. 
\\ \hline
\textbf{Taxometric \newline architecture} & What logical structure defines the target variable? (e.g. binary vs. multi-label vs. continuous) & \textbf{Categorical reductionism \newline vs. inferential density} 
& Pseudo-precision of intensity scores; foreclosure of subject ambiguity and narrative flux. 
\\ \hline
\textbf{Taxonomic \newline target \newline definition} & Which target categories are selected? (e.g. 'crying' vs. 'sad' vs. 'negative emotion') & \textbf{Normative imposition vs. \newline lived ambiguity} 
& Epistemic injustice; stigmatization of caregiving via moralized labels; objectification of the body; narrative fixing into target roles.
\\ \hline
\newpage
\multicolumn{4}{|l|}{\cellcolor[gray]{0.9}\textbf{STAGE C: INFERENCE \& EVALUATION}} \\ \hline
 \textbf{Infrastructure} & Where is inference performed and who retains control? (e.g. AWS, local servers) & \textbf{Analytical power vs. \newline data sovereignty}
 & Secondary data use by third parties; loss of data sovereignty; exposure of unconsenting subjects to proprietary gazes and model training. 
 \\ \hline

\textbf{Implementation \& selection} & Which (pre-trained) models are used? (e.g. off-the-shelf, finetuned, task-specific) & \textbf{Technical efficiency vs. \newline representational accuracy} 
& Misclassification of non-normative or marginalized groups; demographic bias rendering subjects computationally suspect. 
\\ \hline

\textbf{Stochasticity} & How much non-deterministic variance is permitted?  (e.g. temperature of LLMs) & \textbf{Computational flexibility \newline vs. stochastic vulnerability} 
& Inconsistent social verdicts; loss of narrative stability; ``hallucinated" emotional states. 
\\ \hline

\textbf{Thresholding} & At what confidence level is a detection ``validated"?  (e.g. class with highest probability vs. higher than \textit{0.9})& \textbf{Evidentiary breadth vs. \newline pathologizing noise} 
& Probabilistic vulnerability; the ``dial of suspicion" as final arbiter of moral standing; pathologization of ordinary caregiving. 
\\ \hline

\textbf{Performance validation} & How is performance measured and errors interpreted? (e.g. accuracy, recall, manual error analysis) & \textbf{Technical optimization vs. \newline representational justice}
& The ``validation trap"; reinforcement of socioeconomic stereotypes; displacement of lived experience by algorithmic notions of certainty. 
\\ \hline

\multicolumn{4}{|l|}{\cellcolor[gray]{0.9}\textbf{STAGE D: DISSEMINATION}} \\ \hline
\textbf{Narrative \newline framing} & How are outputs synthesized into a coherent story? (e.g. interpretative openness or closure) & \textbf{Analytical humility vs. \newline rhetorical certainty} 
& Narrative fixing; stabilization of children within roles of exploitation; triggering of unmerited state or social intervention. \\ \hline

\textbf{Visualization} & How is visual evidence selected for publication? (e.g. inclusion or exclusion of screenshots) & \textbf{Evidentiary impact vs. \newline cumulative exposure} 
& Spectacularization of harm; emotional hypervisibility; permanent archiving of private domestic spaces (e.g., bedrooms) in the academic record. \\ \hline

\textbf{Publication} & What makes it out of the ``lab"? (e.g. pubblishing raw data, only metadata) & \textbf{Public accountability vs. \newline the right to be forgotten}
& Weaponization by bad actors like vigilante harassment; searchable digital stigma for subjects in their future social or clinical lives. \\ \hline
\end{longtable}
\endgroup

\section{The Co-Production of Vulnerability: A Reflexive Framework for AI Research} 
\label{sec:vdp}

Our analysis inductively identified four mechanisms through which technical decision points amplify or transform the precarity of platformized subjects. The first is \textbf{exposure}, wherein research-led datafication creates a ``secondary life" for platformed content that intensifies risks to intimacy. By assembling frame-level facial archives, for example, researchers transform ephemeral performances into durable, searchable evidence, inviting institutional scrutiny further scaled by the circulation of findings in papers and conferences. This moves subjects from platform niches into a permanent regulatory gaze they cannot exit. Second is \textbf{monetization}, as AI4SG can inadvertently deepen economic instrumentalization when researchers leverage platform APIs or pursue grants and reputational rewards for extracting personal data. In this ``secondary monetization", researchers can mirror the attention-economy extraction they aim to regulate, and further embed subjects' identities within a logic of commercial and academic value. The third mechanism, \textbf{narrative fixing} represents a form of epistemic foreclosure where a subject’s identity is externally scripted and stabilized. Through labeling and interpreting, such as categorizing subjects as ``sad" or as ``exploited", researchers may deny subjects the right to emotional flux or interpretative self-determination, locking them into rigid, research-imposed roles that may contradict their lived experience or future self-conception. By freezing transient expressions into permanent data points, researchers exert a form of representational control that ``fixes" the past as a prescriptive map for the subject's future. Finally, \textbf{algorithmic optimization} precarizes vulnerability by reshaping behaviors to align with technological affordances. By defining protection through specific metrics, data-based research can reinforce a logic where subjects are recognized only if they are machine-readable. This can create a form of ``recursive" vulnerability: for example, families may coach their children to exaggerate smiles and other displays of happiness in order to generate positive affective computing scores from CV models. Research-defined metrics thus become part of the optimization scripts that dictate how platformized lives are performed.

To assist researchers in navigating the protection paradox, we suggest four core reflexive questions:

\begin{enumerate}
    \item \textbf{Are we increasing exposure?} 
    Do our datasets, models, visualizations, or publications inadvertently amplify the visibility, reuse, or circulation of precarious lives?

    \item \textbf{Are we fixing narratives?} 
    Do our labels, proxies, or analytic categories freeze a subject's identity or experience into something imposed, normative, moralized, or misleading?

    \item \textbf{Are we enabling or benefiting from monetization?}
    Who gains value---academic, economic, reputational---from the data, and who bears the risks? What forms of consent or agency are (not) possible?

    \item \textbf{Are we driving algorithmic optimization?}
    Do our metrics, benchmarks, or proxies encourage subjects to become legible to systems designed by us, reinforcing the very logics we aim to critique or resist?
\end{enumerate}

These four questions cannot be addressed in isolation, nor decontextualised from the scientific, cultural, and socio-economic practices in which data-driven activities are performed. Additionally, this reflective exercise cannot be translated into a numerical trade-off. The concern each question raises individually merits attention, and whether alone or in combination with others,  may weigh significantly on the overarching question of whether, and if so, under what conditions, research may be responsibly performed. 

\subsection{A Reflexive Protocol for Data Practice}
Moving from compliance to a form of situated, reflexive ethics requires pausing at key technical decision points. We propose a \textbf{reflexive protocol for data scientists} (\textbf{Appendix \ref{appendix}}) as a practical scaffold designed to help researchers identify the specific junctures where technical choices necessitate pause. The protocol’s primary function is twofold: first, it identifies key junctures within the pipeline; and second, it provides initial questions to interrogate the ethical tensions hidden within that juncture. 
At each stage, the protocol prompts researchers to look beneath the immediate technical question to trace the underlying dynamics of the precarization of vulnerability. 
For example, where a data scientist asks about scale in dataset design (\textit{How many data subjects?}), the protocol triggers a pause to ask: \textit{Does increasing scale primarily diffuse subject-level fixation, or does it instead expand the reach of population-level screening? Which forms of harm—cumulative exposure and monetization, interpretive overreach—are amplified or muted at different scales?} While the protocol offers suggested trade-offs and concrete steps, its core contribution is this interrogative opening. It forces an audit of the four mechanisms—exposure, monetization, narrative fixing, and algorithmic optimization—and requires researchers to consider how ecosystem actors, such as journalists and regulators, might ``lock in" inferred narratives. By identifying these critical junctures and providing questions to navigate them, the protocol shifts data science from a linear execution of tasks into a reflexive, situated practice.

\section{Institutionalizing Reflexivity: EU Law and Research Ethics in the face of AI}
\label{sec:law}

The reflexive junctures identified in our protocol are not merely aspirational; they are increasingly mirrored in the substantive requirements of European data protection and AI governance. Key legal tenets, such as the fairness principle, purpose limitation, and data protection by design, encourage a reflexive and vulnerability-aware methodology rather than a simple compliance checklist. 

EU data protection standards, and the General Data Protection Regulation (GDPR) in particular \cite{EuropeanParliament_CounciloftheEuropeanUnion_2016}, hold instrumental value for protecting inherently vulnerable data subjects.  Critically, the fairness principle mandates that processing of personal data should not result in the unlawful discrimination or exploitation of data subjects or be otherwise unjustifiably detrimental to them \cite[para 69]{EuropeanDataProtectioBoard2020design}. Crucial for realizing this imperative is the implementation of technical and organizational measures to counter asymmetries in power \cite{EuropeanDataProtectioBoard2020design}. 

Power imbalances are acute in research that involves intrusive technologies. From a legal perspective, particular attention is paid to situations in which ``special categories of data" will be processed (Article 9(1) GDPR).  The latter categories include data that may reveal, directly or indirectly (by means of an intellectual operation involving collation or deduction), information about people's racial or ethnic origin, their political and religious beliefs, genetics, health, sex life, or sexual orientation.\footnote{CJEU Case C-21/23 ND v DR, ECLI:EU:C:2024:846, 2024, para 83.}  Biometric data for the purpose of uniquely identifying a natural person is considered sensitive, too. The processing of special categories of data is in principle prohibited as the information concerned risks further precarizing the vulnerability of data subjects to social, economic, or political abuse \cite{Georgieva_Kuner_2020special}. In this regard, special categories of data share a kinship with protected identity traits found in non-discrimination law. In essence, both frameworks aim to counterbalance power as a source of social inequality. That said, research may qualify as an exception insofar as adequate protective measures are in place to safeguard the fundamental rights and interests of data subjects (Article 9(2) (j) GDPR). Additionally, the European Data Protection Board has identified that the processing of sensitive data—a more general notion that includes categories of data beyond those listed in Article 9 of the GDPR, such as location data or financial information—concerning vulnerable data subjects through innovative technological solutions constitutes a high-risk scenario under the data protection impact assessment provision (Article 35 GDPR), warranting a continuous review and reassessment of the envisaged activities, their risks to data subjects, and the safeguards put in place to protect them \cite{Article292017guidelines}.  

Given the substantive ambitions of EU data protection legislation, there is a strong mandate for data-driven research to incorporate reflexive practices as appropriate safeguards. While power imbalances characterize most research settings, a relational conception of vulnerability may help further delineate the safeguards that best address the needs and interests of data subjects in a given situation. In this regard, the principle of data fairness has been interpreted as comprising two protective steps \cite{clifford2018data, malgieri2025scalable, Clifford_2024, Naudts_N.D.}. On the one hand, data subjects find protection against abuse of power through a series of default procedural and legal safeguards, such as transparency and data minimization. On the other hand, processing operations must be preceded by a multi-layered exercise that weighs the purported benefits of the envisaged activities against the negative impact they may have at the individual, collective, and societal levels. 

Applied to data-intensive research, this test would require researchers to assess, among others, whether the research goals are legitimate and the data-driven technologies relied on to realize that purpose are necessary. Under this step, researchers should also consider the safeguards they want to implement and the availability of less intrusive alternatives. Finally, even if the envisaged operations would appear appropriate and necessary, they must not cause disproportionate disadvantage. Rather, a fair balance must be struck between interests involved.\footnote{See also Advocate General Kokott's Opinion to CJEU Case C-157/15 Samira Achbita and Centrum voor Gelijkheid van Kansen en voor Racismebestrijding v. G4S Secure Solutions NV, ECLI:EU:C:2016:382, 2016, para 112.} The outcome might favor abandoning a particular research proposal or using a particular technology in research. The reflective data practices we endorse throughout this paper, and the protocol we offer, help operationalise the delicate balancing exercise envisaged by the law. Moreover, under this view, reflexivity may inform how formal constraints are shaped; for instance, the requirement of transparency is purposeful in equalizing power asymmetries only when information is catered to the needs and interests of the target audience.

\section{Limitations}
\label{sec:limitations}

This paper offers a qualitative, conceptual analysis grounded in a single, empirical case study. While the family vlog investigation provides a concrete anchor for examining how protective intentions can entail data practices that contribute to the precarization of vulnerability, it does not claim empirical exhaustiveness or representativeness. The analysis foregrounds decision points across the AI research pipeline rather than technical evaluations of specific models or quantitative performance; it therefore does not assess accuracy, bias metrics, or error rates, but focuses on how methodological choices structure ethical exposure and vulnerability. This necessarily abstracts from implementation-level variation in order to surface recurring patterns of harm production. The analysis is also geographically and legally situated. Its normative focus centers on EU law, particularly the GDPR. Although these frameworks increasingly influence global AI governance, their interpretation and enforcement do not generalize across jurisdictions. Moreover, national differences within the EU—regarding scientific freedom, institutional autonomy, and research obligations—condition how legal standards are applied and are not exhaustively addressed here. More broadly, the paper’s regulatory framing is constrained by the transnational nature of platformed lives and AI research. Content circulates globally, and influencers increasingly relocate to jurisdictions with divergent regulatory regimes, complicating questions of jurisdiction and accountability. Finally, the analysis focuses on formal research contexts—journalistic, academic, and regulatory—rather than informal, commercial, or platform-internal uses of similar techniques. While related international developments, including recent UN reports, are acknowledged, they are not analyzed in depth.

\section{Conclusion}
\label{sec:conclusion}

Neither dismissing all AI4SG projects as outright technosolutionist, nor uncritically embracing them as a vehicle for accountability, we instead argue for a program of reflexive practice. This program treats research itself as world-making work and demands methodological structures that can hold open the question of how far visibility becomes protective or predatory in different contexts and given moments. In practical terms, we propose a reflexive ethics protocol that offers researchers concrete guidance for working in highly sensitive, platformized environments. The protocol is organized around four recurring decision points in the research pipeline: dataset design (what is collected, from where, at what scale); operationalization (how questions and concepts are rendered as computational objects); inference (which probabilistic outputs are converted into definitive ``evidence", and how); and dissemination and circulation (what leaves the ``lab'', in what form, and for whom). At each point, we surface the ways in which well-intentioned work can slide into renewed extraction, renewed exposure, and renewed authority and interpretative imposition over other people’s lives.

What emerges through our shift of focus, from vulnerable data subjects to vulnerabilizing data practices, is not an already vulnerable subject waiting to be helped, but a set of socio-technical arrangements that actively manufacture the precarization of vulnerability. Beginning from this relational framing allows us to approach the journalist’s request differently. The question shifts from “how can we operate tools that count children in vlogs so policymakers get a ground to act on,” to “what kinds of social and technical worlds are enacted when we do so, which tensions arise, and how can we navigate them?” Data science and AI-powered technologies are not neutral instruments waiting to be applied to a problem. Each carries a set of commitments about what counts as a unit of analysis, what kinds of signals are salient, what categories are real enough to be measured, and what forms of visibility are ``worth" producing. Each encodes a world in which certain bodies, affective states and information can be detected, categorized, and governed, and in which certain actors are authorized to do that categorizing. In this sense, the turn to AI in the name of protection does not merely reflect harms that already exist on platforms. It participates in making a version of those harms actionable to institutions. 
In doing so, it also helps define which people become governable in ways that exceed personal control, which platformized practices become legible as abuse, which platform business models can be framed as exploitative, and which subjects they put at risk of being harmed. 

\section{Endmatter}

\subsection{Ethical Considerations Statement}

Given the normative and procedural orientation of this work, we acknowledge a salient risk of adverse or unintended impact: the presented framework could be selectively adopted, misread, or cited as a form of \textit{ethics washing}. By \textit{ethics washing}, we mean the practice of invoking ethical research primarily to signal compliance or responsibility, e.g. by citing “ethics” sections, checklists, or protocols, without substantively engaging with the specific ethical–political challenges identified, or acting to mitigate them. Substantive engagement, as we use the term, requires \textit{situated} reflection: attention to the particulars of the case at hand (including stakeholders, power relations, data provenance, deployment context, and foreseeable downstream effects), rather than reliance on generic or “off-the-shelf” solutions.
We further recognize that the reflective exercise and tabular artifact proposed in this paper may be misconstrued as exhaustive, treated as a procedural substitute for accountability, or applied in ways that fail to advance the contribution’s intended aim; namely, strengthening the voice and choice of vulnerable or otherwise impacted data subjects. Accordingly, we emphasize that the protocol is not intended as a completeness guarantee. Future applications may require tailored questions, methods, and forms of reflexive inquiry beyond those enumerated here, particularly in contexts characterized by high asymmetries of power, limited contestability, or unclear avenues for redress.
Finally, because the framework necessarily reflects our own interpretations and normative commitments, we invite critical scrutiny and contestation of its assumptions, boundaries, and effects. We encourage researchers, practitioners, and policymakers to assess whether and how the proposed approach meaningfully alters research practice, governance, and decision-making in fair machine learning, and to adapt or reject elements that do not improve protections, participation, or accountability in their specific context.

\subsection{Positionality Statement}

Our perspective is shaped by our positionality as Europe-based researchers writing from relatively privileged positions with access to institutional resources—such as ethics officers and review boards—that allow time for reflection and ethical deliberation. We recognize that this context is specific and that our insights may not fully translate to settings with different social, economic, or institutional conditions. One author is a Latin-American scholar, recently naturalized in Europe and working in academia in the Netherlands. Their training in computer science and digital humanities informs how they attend to identity, toxicity, and vulnerability in datafied environments, while their engagement with decolonial perspectives shapes their interpretation of power relations in computational practices. Another author, a European citizen working in academia in the Netherlands and residing in Germany with two children, brings a perspective grounded in continental philosophical traditions. This formation informs both their research approach and the selection of works cited. A third author, a European citizen working in the retail sector in the Netherlands, draws on media studies to focus on the politics of visibility in the development and use of computer vision technologies. Another author, a European citizen working in academia in the Netherlands, is an interdisciplinary scholar in fundamental rights and information law, with work rooted in political philosophy and legal theory.

\subsection{Generative AI Disclosure Statement}

Generative AI was used to assist with formatting (especially with LaTeX formatting of tables), and reflexively to assist in refining the grammar and fluency of the manuscript. The authors take full responsibility for the text.

\subsection{Acknowledgements}

D.S. Martinez Pandiani, P. Helm, and E. Streefkerk thank the University of Amsterdam's Institute for Advanced Study (IAS) and Data Science Center (DSC) for their financial support through the Joint IAS–DSC Fellowship Programme. L. Naudts was supported by the ``AI, Media \& Democracy Lab - Dutch Research Council project number: NWA.1332.20.009" and the Dutch Journalism Fund (SVDJ) ``Thematische Onderzoeksregeling 2024 - 2026: Oproep AI en nieuwsbehoeften."

\bibliographystyle{ACM-Reference-Format}
\bibliography{sample-base}
\appendix

\section{Appendix}
\label{appendix}

\newcolumntype{L}{>{\raggedright\arraybackslash}X}
\begin{landscape}
\begin{footnotesize} 
\begin{xltabular}{\linewidth}{p{2.7cm} p{3.8cm} p{5.1cm} p{3cm} p{3.6cm}}
    \caption{Reflexive Ethics Protocol for AI Research} \label{tab:protocol} \\
    
    \toprule
    \textbf{Technical \newline Decision \& Question} & \textbf{Abstracted Tensions} & \textbf{Reflexive Protocol Questions} & \textbf{Potential \newline Vulnerabilization(s)} & \textbf{Reflexive Response} \\ \midrule
    \endfirsthead
    
    \multicolumn{5}{c}{{\bfseries \tablename\ \thetable{} -- continued from previous page}} \\
    \toprule
    \textbf{Technical \newline Decision \& Question} & \textbf{Abstracted Tensions} & \textbf{Reflexive Protocol Questions} & \textbf{Potential \newline Vulnerabilization(s)} & \textbf{Reflexive Response} \\ \midrule
    \endhead

    \hline \multicolumn{5}{r}{{Continued on next page...}} \\
    \endfoot

    \bottomrule
    \endlastfoot

    \rowcolor{gray!20}
    \multicolumn{5}{l}{\textbf{STAGE A: DATASET DESIGN}} \\ \midrule
    
    \textbf{Scale}: \newline How many data subjects are included in the dataset? & \textbf{Subject-specific fixation vs. population-level scrutiny.} Narrow scale (e.g., one channel) concentrates interpretive attention and cumulative exposure on specific subjects. Broad scale diffuses this fixation but expands the total population subjected to automated screening. & How does dataset scale interact with the kind of inference being performed? Does increasing scale primarily diffuse subject-level fixation, or does it instead expand the reach of classificatory or inferential scrutiny? Which forms of harm (cumulative exposure, interpretive overreach, population-level screening) are amplified or muted at different scales? & Exposure through hyperfocused inclusion; intensified scrutiny of specific subjects; increased traceability and narrative fixing over time. & Explicitly document the rationale for dataset scale. Weigh depth versus breadth as an ethical choice. Consider hybrid strategies (partial aggregation, temporal subsampling). Ask if aggregation can reduce fixation without undermining evidentiary aims. \\ \midrule
    
    \textbf{Selection criteria}: \newline Based on what criteria are data subjects selected? & \textbf{Attention mirroring vs. structural opacity.} Selecting based on high visibility reinforces platform ``exposure" logics. Conversely, seemingly neutral criteria may obscure how platforms systematically elevate particular forms of vulnerability or labor. & Why these data subjects rather than others? What logics of relevance, harm, or impact are embedded in the selection criteria? Does the sampling strategy amplify visibility for subjects already shaped by platform attention, or does it obscure structural mechanisms by privileging apparent neutrality? & Exposure through inclusion; repeated analyzability and inference; intensified scrutiny; representational fixation where specific subjects become emblematic cases. & Treat selection as an explicit ethical decision point. Compare alternative sampling strategies (controversial, engagement-based, random) in terms of yield and vulnerability. Make criteria explicit and open to contestation. \\ \midrule
    
    \textbf{Comparability \& \newline aggregation}: \newline Under what assumptions are subjects treated as aggregable? & \textbf{Representational flattening vs. over-individualization.} Aggregation enables generalization but erases the context-specific power dynamics of different subjects/channels. Disaggregation preserves context but risks isolating systemic harm as merely idiosyncratic or personal. & Can the selected subjects be meaningfully compared and aggregated? Under what assumptions are subjects treated as aggregable? Does selecting these subjects reinforce historical narratives of representation? What differences are treated as analytically irrelevant? Does aggregation assume equal baseline risk, thereby intensifying scrutiny of some bodies? & Representational flattening that obscures differential vulnerability; normalization of certain expressions as baseline cases; misinterpretation of context-dependent signals as generalizable patterns. & Make aggregation assumptions explicit. Conduct positionality-aware audits of comparability. Use partial aggregation strategies. Retain the ability to disaggregate results when power or exposure meanings vary. \\ \midrule
    
    \textbf{Resolution \& \newline granularity}: \newline At what level of granularity are data subjects rendered analyzable? & \textbf{Analytical sensitivity vs. inferential density.} Higher resolution (frame-by-frame) increases evidentiary precision but multiplies the instances of bodily inspection and (e.g. emotional) profiling a subject undergoes. & What is the minimum granularity required to answer the question credibly? Does increased granularity substantively change the claim, or merely increase precision? Do different tasks justify different granularities? Is frame-level analysis necessary to demonstrate harm, or would coarser units preserve the claim while reducing traceability? & Hyper-granular exposure through repeated frame-level analysis; intensified inferential density per subject; increased traceability and narrative closure driven by fine-grained temporal data. & Default to the least intrusive granularity. Justify escalations task-specifically. Use staged designs (thumbnails first, frames only if needed). Document what is lost and gained at each level. \\ \midrule
    
    \textbf{Access \& reuse \newline justification}: \newline On what grounds is data deemed eligible for inclusion? &  \textbf{Legal accessibility vs. ethical permissibility.} Treating public availability as a proxy for consent ignores how research-led reuse extends data life and produces durable, secondary records (e.g., inferred emotional states) that subjects cannot exit. & What assumptions link public accessibility to ethical reuse? Is monetization treated as a necessity or a normative filter? How does inclusion extend the reach of exposure? Does reuse violate emotional privacy by creating durable records of internal states beyond original context? & Extended data life through reuse and recontextualization; amplification of exposure; normalization of platform-defined publicness as ethical consent. & Distinguish legal access from ethical justification. Minimize retention, duplication, and redistribution. Document why inclusion is necessary for the claim. Revisit boundaries when analytic goals shift. 
    \\ \midrule

    \rowcolor{gray!20}
    \multicolumn{5}{l}{\textbf{STAGE B: OPERATIONALIZATION}} \\ \midrule
    
    \textbf{Unit of analysis \& \newline localization}: \newline  At what resolution is the analysis performed? & \textbf{Hyper-identification vs. contextual blur.} Individual localization provides ``proof" for granular claims but necessitates identifiable biometric archives. Scene-level detection protects identity but may obscure the specific context of participation. & How does the chosen unit mediate the visibility of power? Does the unit differentiate between incidental presence and active participation? At what threshold does the pursuit of evidentiary precision transition into persistent biometric surveillance? & Persistent individual tracking; creation of identifiable biometric corpuses; individualization of systemic harms; mischaracterization of incidental life. & Explicitly justify the need for localized tracking over global tags. Evaluate if anonymized proxies (pose/silhouette) can replace biometrics. Audit whether individualization is necessary for the claim. \\ \midrule
    
    \textbf{Taxometric \newline architecture}: \newline What logical structure defines the target variable? & \textbf{Categorical reductionism vs. inferential density.} Simple structures flatten experience; complex/continuous structures preserve ``nuance" but subject the person to a ``thickness" of constant, overlapping automated judgment. & Does the mathematical structure (e.g., continuous 0–1 scales) create a ``thickness" of constant inspection? Does the architecture force a singular, stable ``state" onto a subject, or allow for narrative flux and performative ambiguity? & Pseudo-precision of intensity scores; normalization of constant automated scrutiny; foreclosure of subject ambiguity; ``state-forcing" that denies change. & Document the rationale for intensity scales. Avoid multi-labeling for ambiguous behaviors where ``mixed" signals lead to automated pathologization. Incorporate ``unknown" categories. \\ \midrule
    
    \textbf{Taxonomic \newline target definition}: \newline Which target categories are selected? & \textbf{Normative imposition vs. lived ambiguity.} Imposing universalist (WEIRD) taxonomies overwrites culturally specific, neurodivergent, or performative expressions with a standardized, adult-centric or colonial gaze. & Which cultural, colonial, or normative assumptions are embedded in the ground truths? When, where, and by whom were the used taxonomies developed, and under what power dynamics? What systems of knowledge and identity categories are included and excluded? Are we committing epistemic injustice by overdetermining or overwriting subjects' right to define themselves? How do these definitions contribute to stigmatization or dehumanization? Does the label facilitate invasive domestic cataloging? & Epistemic injustice; ``narrative fixing" via moralized labels; objectification of the body; stigmatization of ordinary caregiving or cultural norms. & Scrutinize labels for Western/adult/class-based bias. Treat thresholds as socially constructed. Explicitly weigh the risk of ``invasive cataloging" against evidentiary necessity. 
    \\ \midrule

    \rowcolor{gray!20}
    \multicolumn{5}{l}{\textbf{STAGE C: INFERENCE AND EVALUATION}} \\ \midrule
    
    \textbf{Infrastructure}: \newline  Where is the computation performed and who retains control? & \textbf{Analytical power vs. data sovereignty.} Utilizing hosted models (Big Tech APIs) can provide state-of-the-art reasoning but can lead to the leakage of intimate (sensitive) data into commercial ecosystems. & Does the use of hosted models violate the data sovereignty of the subject? Is the evidentiary gain worth the secondary exposure? Am I sending unconsented imagery to a third-party server where it may be used for proprietary model training? & Secondary data use by third parties; loss of data sovereignty; exposure of unconsenting subjects to proprietary gazes. & Explicitly justify the use of cloud APIs over local inference. Use anonymization or obfuscation techniques before transmission. Prioritize local, open-weights models where feasible to retain sovereignty. \\ \midrule
    
    \textbf{Implementation \& \newline selection}: \newline Which pre-trained models are used? & \textbf{Technical efficiency vs. representational accuracy.} Relying on off-the-shelf, (adult-, white-centric encoders) can scale systemic bias, misinterpreting idiosyncratic behaviors through a narrow representational sieve. & Is the model representative of the demographic and cultural context? What systemic biases are being ``imported" via pre-training? Does an adult-centric (e.g. nudity or emotion) detector ignore the specific developmental baselines of minors? & Mass misclassification of marginalized groups; ``visibility distortion" where demographic bias renders subjects computationally suspect or illegible. & Audit models for demographic bias. Fine-tune on representative datasets to correct adult-centric skew. Document representational limits and avoid general-purpose models for sensitive niche populations. \\ \midrule
    
    \textbf{Stochasticity}: \newline How much non-deterministic variance is permitted? & \textbf{Computational flexibility vs. stochastic vulnerability.} Non-deterministic reasoning (LLMs) can produce inconsistent ``truths" based on prompt phrasing or temperature settings. & How does non-deterministic variance affect the reliability of the research claim? Is the model's interpretation stable across different runs? Does an interpretation of ``distress" vary based on the arbitrary phrasing of a prompt? & Inconsistent social verdicts; loss of narrative stability; ``hallucinated" emotional states; lack of replicable evidentiary grounds. & Use low ``temperature" settings to minimize variance. Perform multi-run consistency checks. Standardize prompting logic and report the variance in model interpretations to avoid over-claiming certainty. \\ \midrule
    
    \textbf{Thresholding}: \newline At what confidence level is a detection ``validated"? & \textbf{Evidentiary breadth vs. pathologizing noise.} High thresholds protect from over-surveillance but risk regulatory abandonment; low thresholds may prioritize protection at the cost of framing often ordinary and domestic life as suspect. & How do thresholds mediate the production of ``truth"? Does the chosen threshold ``erase" signs of discomfort to avoid noise, or does it ``flag" ordinary caregiving (e.g., bathing) as a regulatory anomaly? & Probabilistic vulnerability; regulatory abandonment (false negatives) vs. pathologization of ordinary care (false positives). & Perform sensitivity analysis on thresholds. Justify probability cut-offs in relation to the specific risk of harm. Incorporate human-in-the-loop review for detections near the threshold boundary. \\ \midrule
    
    \textbf{Performance \& \newline validation}: \newline How is performance measured and errors interpreted? & \textbf{Technical optimization vs. representational justice.} Aggregate metrics (F1/Accuracy) can mask disparate impacts on vulnerable subsets and stabilize problematic affective narratives. & Do aggregate metrics mask disparate impacts on social groups? Could errors reinforce existing political or socioeconomic narratives? Are errors concentrated in families with specific socioeconomic attributes? & The ``validation trap"; stabilization of problematic affective narratives; reinforcement of socioeconomic stereotypes; displacement of lived experience. & Conduct disaggregated performance audits (e.g., by race, age, class). Prioritize error-type analysis (false positives vs. negatives) over aggregate scores. Situate performance metrics within cultural and contextual limits. \\ \midrule

    \rowcolor{gray!20}
    \multicolumn{5}{l}{\textbf{STAGE D: DISSEMINATION}} \\ \midrule
    
    \textbf{Narrative framing}: \newline How are complex outputs synthesized into a coherent story? & \textbf{Analytical humility vs. rhetorical certainty.} Translating probabilistic model outputs into definitive moral or legal narratives can create an interpretive closure that overrides subject agency. & Does the narrative allow for interpretive pluralism, or does it suggest closure and certainty? Am I framing an ``80\% confidence" distress score as a definitive case of neglect? What moral connotations do the results imply for the subjects? & Narrative fixing; stabilization of subjects within roles of exploitation; triggering of unmerited social or state intervention; erasure of performative context. & Adopt a posture of analytical humility. Explicitly report uncertainty and confidence intervals. Use qualifying language that acknowledges the limitations of model-ascription over lived experience. Allow for alternative interpretations in the discussion. \\ \midrule
    
    \textbf{Visualization}: \newline How is visual evidence selected for public or academic inquiry? & \textbf{Evidentiary impact vs. cumulative exposure.} The pursuit of ``visual proof" for peer or public audiences can normalize the re-exposure of subjects' (private) lives and replicates the platform's logic of engagement. & Does the inclusion of visual evidence replicate the ``spectacle" of harm? Am I including a frame of a subject's bedroom that permanently archives their private space in the academic record? Is there a non-visual way to convey the evidentiary weight? & Spectacularization of harm; emotional hypervisibility; permanent archiving of private domesticity; commodification of subjects' inner lives. & Prioritize non-visual evidence (e.g., aggregate data, synthetic examples). If visuals are necessary, use aggressive obfuscation (beyond simple blurring). Evaluate if the visual ``proof" provides a utility that outweighs the harm of permanent re-exposure. \\ \midrule
    
    \textbf{Publication}: \newline What makes it out of the ``lab"? & \textbf{Public accountability vs. the right to be forgotten.} Publishing datafied evidence creates a secondary life for the data, resulting in potentially permanent digital footprints that subjects may not consent to nor be able to exit. & Does naming specific channels provide a roadmap for harassment or ``vigilante" intervention? Could publication of these findings be used against the data subject in their future social or clinical life? How might findings be weaponized? & Weaponization by bad actors; ``hit-list" or roadmap creation for harassment; permanent digital branding; searchable digital stigma for subjects as they age. & Redact specific identifiers (channel names, handles) unless absolutely necessary for public accountability. Anticipate downstream weaponization and include explicit disclaimers regarding the limits of the data. Weigh the pursuit of systemic reform against the risk of creating a permanent, searchable record of a subject's vulnerability. \\ 
    
\end{xltabular}
\end{footnotesize}
\end{landscape}

\clearpage
\twocolumn
\end{document}